\documentclass[aps,prb,reprint,amsmath,amssymb,floatfix,superscriptaddress,longbibliography]{revtex4-1}
\usepackage{amssymb}
\usepackage{amsmath}
\usepackage{graphicx,xcolor,multirow,bm,hyperref}
\usepackage[ignoreunlbld,norefs,nocites]{refcheck}
\begin{document}
\title{Electron correlation effects and spin-liquid state in the Herbertsmithite Kagome lattice}
\author{Sam Azadi}
\email{sam.azadi@manchester.ac.uk}
\affiliation{Department of Physics and Astronomy, University of Manchester, Oxford Road, Manchester M13 9PL, United Kingdom}

\author{T. D. K\"{u}hne}
\affiliation{Center for Advanced Systems Understanding, Untermarkt 20, D-02826 G\"orlitz, Germany}

\affiliation{Helmholtz Zentrum Dresden-Rossendorf, Bautzner Landstra{\ss}e 400, D-01328 Dresden, Germany}

\affiliation{%
TU Dresden, Institute of Artificial Intelligence, Chair of Computational System Sciences, N\"othnitzer Stra{\ss}e 46
D-01187 Dresden, Germany
}
\date{\today}
\begin{abstract}
We employ real-space variational quantum Monte Carlo methods with resonating valence bond many-body wave functions to investigate electron correlation effects in the Kagome system $Zn_xCu_{4-x}O_6$. Using three trial wave functions of the Slater-Jastrow type, where (i) only the Jastrow correlation factor is optimized and the orbitals obtained by density functional theory, (ii) both the Jastrow factor and the Slater determinant are optimized, and (iii) additionally the Slater determinant is substituted by an antisymmetrized-geminal power wave function, we analyze static and dynamic correlation energies across concentrations $x=0, \frac{1}{3}, \frac{2}{3}, 1$. Our results show that the correlation energy increases with
the concentration of $Zn_x$. Optimizing the Slater determinant significantly enhances the correlation energy by approximately $\sim -73(4) mHa$ per electron. Eventually, the emergence of a quantum spin liquid state driven by a long-range correlation energy is discussed.
\end{abstract}

\maketitle

\section{Introduction} 
A quantum spin liquid (QSL) is a state of matter that arises in certain magnetic systems. Unlike ordinary magnets, where the magnetic moments (or spins) of electrons align in a regular pattern at low temperatures, in a spin liquid, the spins do not exhibit such an order even as the temperature approaches absolute zero \cite{Norman2016,Broholm2020,Han2012,Chisnell2015}. This lack of long-range magnetic order is due to quantum fluctuations and can occur in systems where the spins are highly frustrated. Magnetic frustration occurs when spins are subject to competing exchange interactions that cannot be satisfied simultaneously, which is usually a result of specific lattice geometries (such as the triangular or kagome lattices) or interactions. For example, in antiferromagnetic systems, where spins prefer to be antiparallel to their neighbors, a triangular arrangement creates a situation where not all spins can be antiparallel to each other, leading to a frustrated configuration \cite{Yan2011,Savary2016,Balents2010}.

The term "liquid" in spin liquid does not refer to a liquid in the conventional sense, but rather the fluid-like disordered state of spins. In such a state, the spins constantly fluctuate and cannot settle into a fixed arrangement. Despite this disorder at the microscopic level, spin liquids can have emergent properties such as a well-defined gauge structure and fractionalized excitations (e.g., spinons, which carry spin but no charge) \cite{Han2011,Helton2010,Fu2015,Banerjee2016,Knolle2014}. Because QSLs can maintain quantum entanglement over a range of temperatures and conditions, they are speculated to have potential applications in quantum computing \cite{Broholm2020}. Their stable entangled state could be used to store and manipulate quantum information with less decoherence, which is a major challenge in the development of quantum computers. There is a strong interest in understanding whether there are any connections between spin liquids and high-temperature superconductivity. Some theories suggest that the mechanism behind high-temperature superconductivity in certain materials may be related to the kind of magnetic interactions seen in spin liquids\cite{PALee2008}.

QSLs can be further classified into different types on the basis of their excitation spectra and whether or not they break time-reversal and other symmetries. They can host exotic excitations such as spinons, which are not possible in other forms of matter. These are fractionalized excitations that carry spin but no charge. However, observing and conclusively demonstrating a spin liquid phase in a material is experimentally challenging and often requires a combination of experimental techniques, including neutron scattering, muon spin rotation, and nuclear magnetic resonance \cite{Helton2007,Mendels2007}.

Originally, the definition of QSL comes from Anderson's seminal paper \cite{Anderson1987}, where he proposed a quantum liquid ground state of an antiferromagnet named a resonance valence bond (RVB) state which is widely used to study the kagome Heisenberg antiferromagnet with a corner-shared triangle structure \cite{Yan2011,Savary2016,Balents2010} and high-temperature superconductivity\cite{AndersonPRL87, Baskaran2003}. The RVB theory describes the electron pairing that can occur in molecules and crystals, particularly those that exhibit strong electronic correlations. Therein, the conventional idea of electrons forming localized bonds (as in a typical covalent bond where electrons are shared between two atoms) is replaced by a picture in which electron pairs are delocalized and shared among many atoms. These electron pairs are termed "valence bonds", and their delocalized nature allows them to resonate over different configurations, giving rise to the idea of "resonance". This is somewhat analogous to the resonance concept in molecular orbital theory, where electrons are delocalized on entire molecules, as originally proposed by Linus Pauling \cite{Pauling} and successfully applied to aromatic compounds containing the benzene ring.

Herbertsmithite, chemically designated as $ZnCu_3(OH)_6Cl_2$, is a mineral of significant interest since it is one of the few known materials that can achieve a QSL state \cite{Freedman2010,Helton2007,Matan2006,Pein2011,Banerjee2017,Han2016,Zorko2017,Pilon2013,Vries2009,Olariu2008}. Its crystal structure features a kagome lattice, a two-dimensional network of corner-sharing triangles formed by copper $Cu$ ions, which carry magnetic moments (spins) due to the unpaired electrons in the $d$ orbitals of copper \cite{Helton2007,Mendels2007}. In herbertsmithite, the kagome lattice geometry results in significant magnetic frustration for the $Cu^{2+}$ spins that prevent them from settling into a simple ordered state as temperature decreases. Instead, these spins remain disordered, leading to a continuum of many-body quantum states that are all equally favorable energetically, an indication of the quantum spin liquid state \cite{Khuntia2020,Barthelemy2022}. Herbertsmithite is thus a prime candidate for experimental and theoretical studies to elucidate the properties of QSLs. 
\begin{figure}
    \centering
    \includegraphics[scale=0.2]{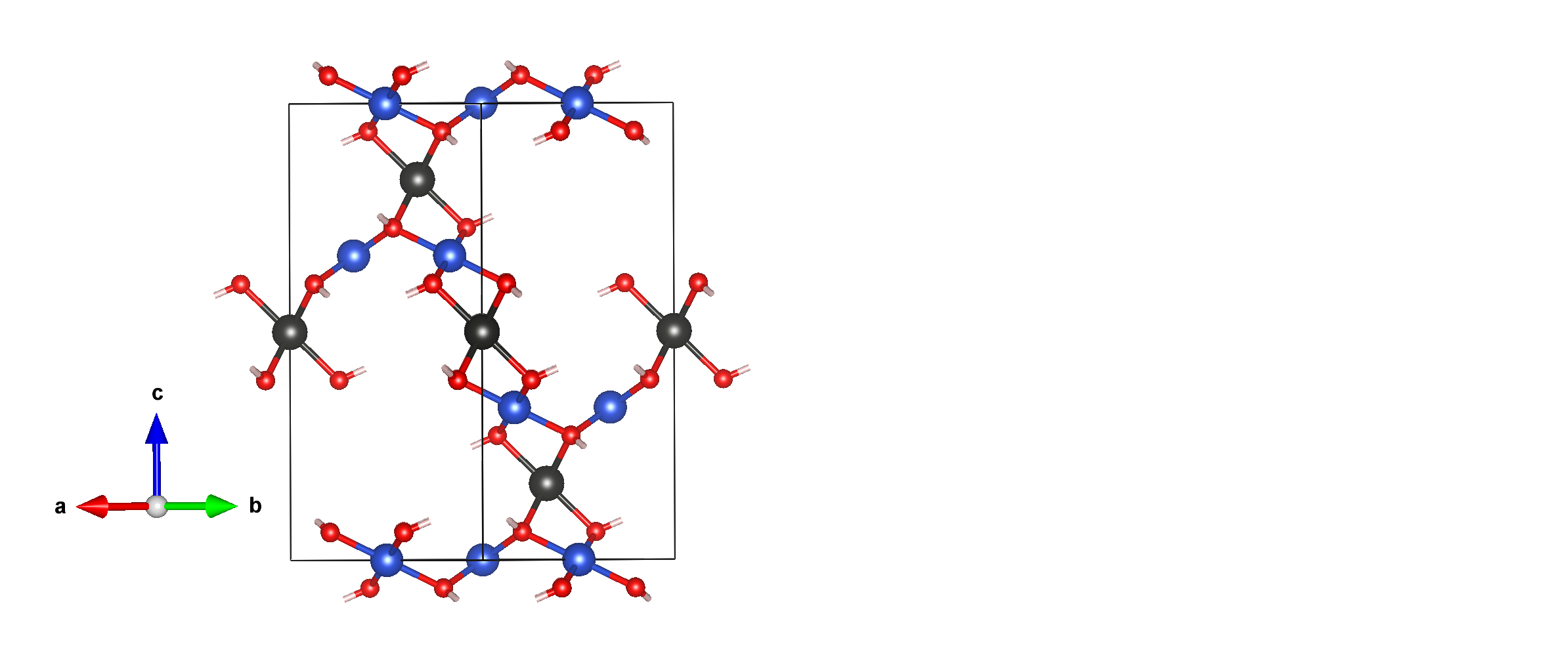}
    \includegraphics[scale=0.12]{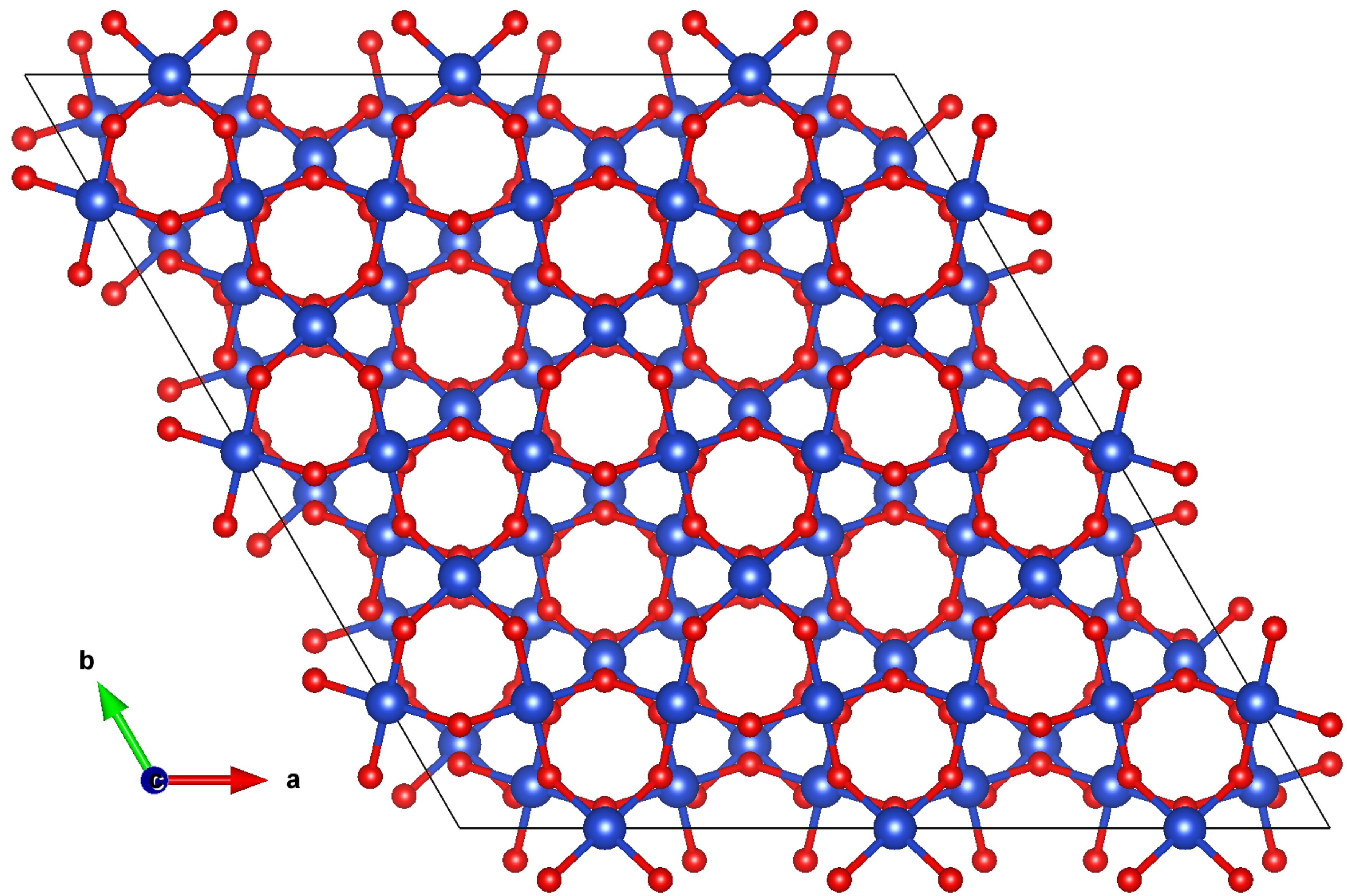}    
    \caption{(Top panel) Crystal structure of Herbertsmithite $ZnCu_3(OH)_6Cl_2$. Zinc atoms are black, copper blue, and oxygen atoms are red. Chlorine and hydrogen atoms are not shown for clarity. There are three sites between $CuO_2$ planes along the $c$ axis which we considered to be occupied by $Cu$ or $Zn$. (Bottom panel) Kagome lattice structure of two-dimensional $CuO_2$ plane. }
    \label{fig:herb}
\end{figure}

In this work, we study the nature of electron correlation and the existence of a quantum spin-liquid state in the $Zn_xCu_{4-x}O_6$ (Fig.\ref{fig:herb}) system as a function of $Zn_x$ concentration $x=0, \frac{1}{3}, \frac{2}{3}, 1$, using the quantum Monte Carlo (QMC) method. We consider three substitution sites between the $CuO_2$ layers that can be occupied by $Cu$ or $Zn$. It was shown that because $Zn^{2+}$ ions are not Jahn-Teller active they prefer the triangular sites over the kagome sites. Neutron powder diffraction measurements on the $x=1$ compound showed that the kagome site is $90\%$ occupied by $Cu^{2+}$ ions \cite{SHLee2007}. In this system with strong local Coulomb interactions, the atomic substitutions affect the local properties of the system.  $ZnCu_3(OH)_6Cl_2$ system was proposed as an ideal candidate for the quantum kagome lattice, where the spins form a two-dimensional triangle of corner sharing \cite{Helton2007}.  The static and dynamic spin correlations as a function of $Zn_x$ concentration $x$ were studied by elastic and inelastic neutron measurements \cite{SHLee2007}, and it was shown that the static and dynamic spin correlations can be altered by doping concentration. Our QMC results indicate that the electron correlation increases by doping with $Zn_x$. 

Our main tool in this work is variational Monte Carlo (VMC), as its results are not affected by the sign problem \cite{Matthew2001}. This method was successfully applied to many systems to calculate electronic and structural properties \cite{Matthew2001,Sorella2011,Marchi2011,Mazzola2014,Devaux2015,Marchi2009,Malatesta}.  For materials without strong electron-electron correlation, including the elements in the first and second row, the fixed node approximation used in diffusion Monte Carlo is less dependent on single-particle orbitals than transition metal compounds, where the electron correlation is much stronger \cite{Kolorenc2010}. Hence, as demonstrated by our results, full optimization of single-particle orbitals using QMC is essential to recover most of the electron correlation. Our results demonstrate that the full optimization of the trial wave function, in which the single-particle orbitals and the Jastrow terms are optimized simultaneously, is the right approach for accurate many-body calculations of systems with possible spin-liquid nature.   

The rest of this paper is organized as follows. Section \ref{QMC} describes the details of our VMC calculations, including the type of trial wave function. Section \ref{results} explains our wave function optimization procedure with results, discusses the static and dynamic correlation energies as a function of $Zn_x$ doping, and investigates the effect of the basis set size for single-particle orbitals on the VMC results. Section \ref{AGPWF} shows the results for the electron pairing energy in the two-dimensional $CuO$ layer, and section \ref{conclude} concludes our results. 
 
\section{Quantum Monte Carlo calculations}\label{QMC}
We deployed QMC methods to find the ground state wave function of the system, which minimizes the total energy. Our simulation cell for system $Zn_xCu_{4-x}O_6$ includes thirty atoms and $336$, $337$, $338$, and $339$ electrons for the concentration $Zn_x$ of $x=0$, $1/3$, $2/3$, and $1$, respectively. The core electrons of the $Zn$, $Cu$, and $O$ atoms were replaced by correlation-consistent pseudopotentials \cite{ccECP1,ccECP2}. The simulation cell is subject to periodic boundary conditions with standard Ewald summations for the long-range Coulomb interaction. All simulations were performed with a real wave function at the $\Gamma$-point and a development version of the TurboRVB code \cite{TurboRVB}. The experimental Herbertsmithite geometry \cite{Norman2016} was used in all of our simulations.

The main ingredient of our QMC calculations is the RVB many-body WF named JAGP \cite{Marchi2009}, as it presents the product of a Jastrow function $J$ and an antisymmetrized geminal power (AGP) determinant part $\Psi_{\text{AGP}}$. The determinant part is written as:
\begin{equation}
    \label{eq:AGP}
    \Psi_{\text{AGP}} (\mathbf{R}) = \mathcal{A} \Pi_{i=1}^{N_{\downarrow}} \phi(\mathbf{r}_{i}^{\uparrow}, \mathbf{r}_{i}^{\downarrow} ) \Pi_{j=1+N_{\downarrow}}^{N_{\uparrow}} \phi_{j}(\mathbf{r}_{j}^{\uparrow}) 
\end{equation}
where $\mathcal{A}$, $\mathbf{R} = \left \{ \mathbf{r}_{1}^{\uparrow}, \cdots, \mathbf{r}_{N_{\uparrow}}^{\uparrow}, \mathbf{r}_{1}^{\downarrow},\cdots, \mathbf{r}_{N_{\downarrow}}^{\downarrow} \right \} $, and $\phi(\mathbf{r}_{i}^{\uparrow}, \mathbf{r}_{i}^{\downarrow}) = \phi(\mathbf{r}_{i}^{\downarrow}, \mathbf{r}_{i}^{\uparrow} )$, are the antisymmetrization operator, the $3N$-dimensional vector of electron coordinates, and a symmetric orbital function describing the singlet pairs, respectively. The last part in Eq.\ref{eq:AGP} $\phi_j(\mathbf{r})$ is the unpaired orbitals we used for systems with $x=1/3,1$. The pairing function was expanded in terms of molecular orbitals (MOs)
\begin{equation}
 \phi(\mathbf{r}^{\uparrow}, \mathbf{r}^{\downarrow}) = \sum_{i=1}^{M} \alpha_i \psi_i^{MO}(\mathbf{r}^{\uparrow}) \psi_i^{MO}(\mathbf{r}^{\downarrow}),
\end{equation}
where the sum is over $M\geq N_{el}/2$ MOs that are expanded in a Gaussian singe-particle basis set ${\chi}$ centered on the atomic position $\psi_i^{MO}(\mathbf{r}) = \sum_j \beta_{ij} \chi_{j}(\mathbf{r})$  \cite{OzoneTurboRVB}. We used an uncontracted Gaussian basis of $10s8p6d2f$ orbitals for $Cu$ and $Zn$, and $4s3p2d$ orbitals for $O$ atom.  The variational parameters in our $AGP$ wave function are $\alpha_i$, $\beta_{ij}$, and the exponents of the uncontracted Gaussian basis set ${\chi}$ with initial values chosen from cc-pCVTZ \cite{ccp} basis set. The condition of $M = N_{el}/2$, where $N_{el}$ is the number of electrons, produces the single Slater determinant $WF$, whereas $M > N_{el}/2$ introduces a static correlation between opposite spin electrons through the pairing function which can be measured as a decrease in the variational ground state energy \cite{Marchi2009,Marchi2011}. In our simulations, the MOs were obtained from density functional calculations \cite{Azadi2010} with the local density approximation (LDA) \cite{lda} using the same uncontracted Gaussian basis set described above. 

The Jastrow term in the trial $WF$ is responsible for the dynamic correlation between two electrons and is conventionally categorised into a homogeneous two-body factor $J_{\text{2b}}$ which is a function of the relative distance between two electrons, and a non-homogeneous terms depending on the position of one, three-body term $J_{\text{3b}}$, or two atoms, four-body term $J_{\text{4b}}$. The one-body function $J_{\text{1b}}$ of Jastrow compensates for change in the single-particle density caused by other Jastrow terms and also satisfies the electron-ion cusp condition. The one- and two-body terms $J_{\text{1b}}$ and $J_{\text{2b}}$ in our $WF$ are defined as 
\begin{equation}
    J_{\text{1b}} = exp 
    \left[ \sum_{ia} -(2Z_a)^{3/4} u(Z_a^{1/4}r_{ia}) + \sum_{ial} g_{l}^{a} \chi_{al}^{J}(\mathbf{r}) \right]
\end{equation}
and
\begin{equation}
    J_{\text{2b}}= exp \left[\sum_{i<j} u(r_{ij}) \right]  
\end{equation}
where $i, j$ are electrons indices and $l$ runs over different single-particle orbitals $\chi_{al}^{J}$ centred on atomic centre $a$. $r_{ia}$ and $r_{ij}$ are electron-ion and electron-electron distances, respectively. The corresponding cusp conditions are fixed by using two different forms for the homogeneous part $u_{F} = r/(2(1+ar))$ \cite{Fahy90} and $u_{C} = \frac{b}{2}(1-exp(-r/b))$ \cite{Ceperley78}, where $r$ is the relative distance between two electrons. $g_l^{a}$, $a$, and $b$ are variational parameters. The three- and four-body Jastrow are given by
\begin{equation}
    J_{\text{3b}}J_{\text{4b}} = exp \left[ \sum_{i<j} f(\mathbf{r}_i, \mathbf{r}_j) \right]
\end{equation}
where $f(\mathbf{r}_i, \mathbf{r}_j)$ is a two-electron coordinate function that is expanded into the same one-particle basis used for $J_{1\text{b}}$
\begin{equation}
    f(\mathbf{r}_i, \mathbf{r}_j) = \sum_{ablm} g_{lm}^{ab} \chi_{al}^{J}(\mathbf{r}_i) \chi_{bm}^{J}(\mathbf{r}_j)
\end{equation}
where $g_{lm}^{ab}$ are optimizable parameters. Three-body electron-ion-electron and four-body electron-ion-electron-ion terms are described by the diagonal matrix elements $g^{aa}$, and off-diagonal $g^{ab}, a\neq b$ matrix elements, respectively. For Jastrow single-particle orbitals, we used uncontracted Gaussian basis sets of $4s3p2d$ for $Cu$, and $Zn$ atoms and $3s2p1d$ for $O$ atom. The complete form of the Jastrow factor $J(\mathbf{R})=J_{1\text{b}}(\mathbf{R})J_{2\text{b}}(\mathbf{R})J_{3\text{b}}(\mathbf{R})J_{3\text{b}}(\mathbf{R})$ used in this work is effective for suppressing higher energy configurations occurring when electrons are too close as the concentration of $Zn_x$ in the system varies. 

We call the initial wave function with $M=N_{el}/2$ and MOs generated by DFT $JDFT\text{-}WF$, where only the Jastrow term was optimized by energy minimization using two techniques of linear basis \cite{Umrigar2007} and stochastic reconfiguration (SR) \cite{Sorella98}. We optimized the Gaussian exponent of the determinant part in $JDFT\text{-}WF$ and called the wave function $JSD\text{-}WF$. Our optimization strategy with the corresponding results is discussed in the next section.  We converted the well-optimized $JSD\text{-}WF$ into $JAGP\text{-}WF$ without losing any information as the maximum overlap between two wave functions is obtained during conversion \cite{TurboRVB}. 

We calculated the electronic correlation as a function of $Zn_x$ concentration in $Zn_xCu_{4-x}O_6$ system using $JDFT$ and $JSD$ wave functions. We also computationally measured the finite-size pairing energy in the two-dimensional $CuO_2$ layer of Herbertsmithite by comparing the energy gain in $JAGP\text{-}WF$ with respect to $JSD\text{-}WF$ . If this energy gain is finite at the thermodynamic limit, it indicates that the $BCS\text{-}WF$ formed by the electron singlet is stable. Due to the large number of electrons in the simulation cell, we were unable to extrapolate our results to the thermodynamic limit to correct the many-body finite size errors \cite{Azadi2015}. All QMC simulations were performed on 40 computational nodes with 128 cores per node and one walker per core. 

\section{Results and discussion} \label{results}
\begin{figure*}[!ht]
    \centering
    \begin{tabular}{ccc}
        \includegraphics[scale=0.38]{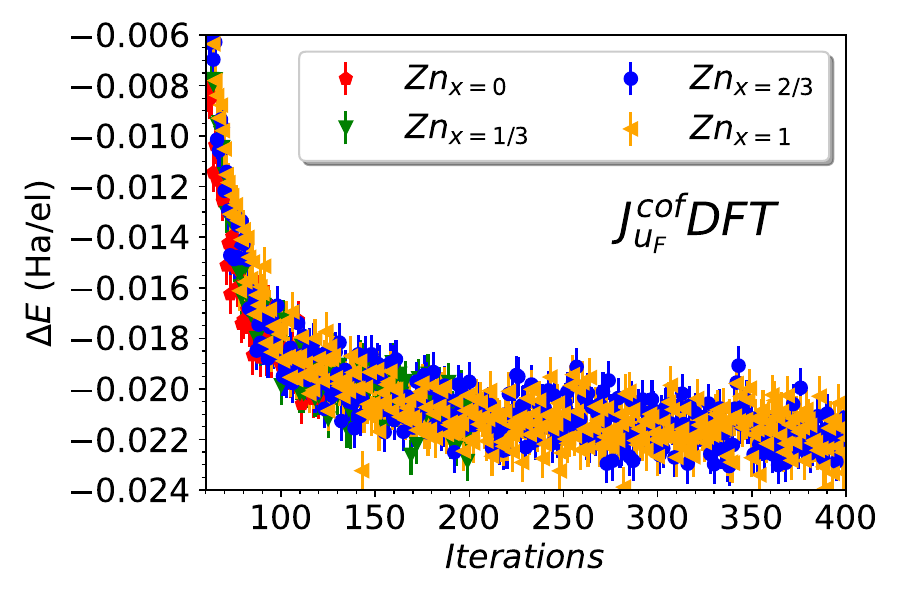}  & 
        \includegraphics[scale=0.38]{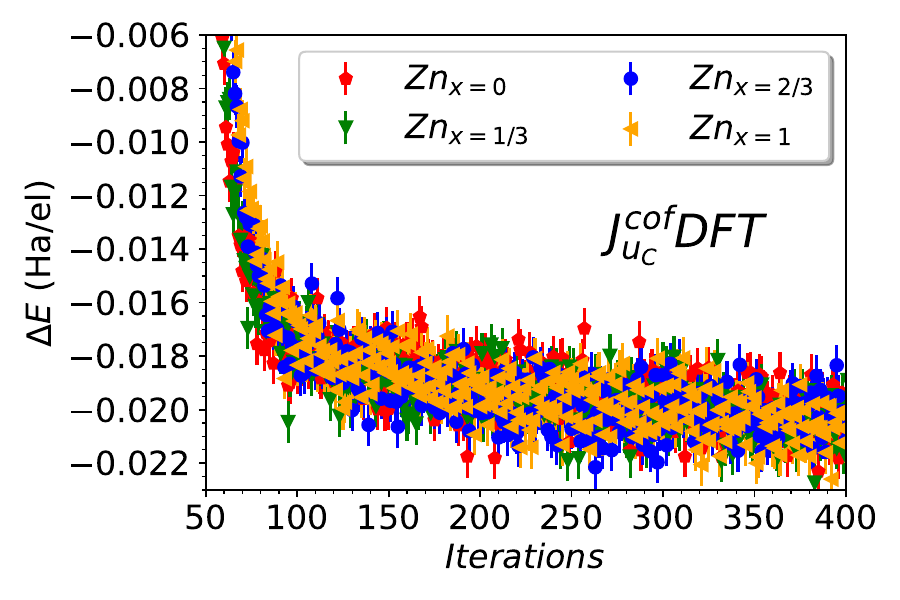} & 
        \includegraphics[scale=0.38]{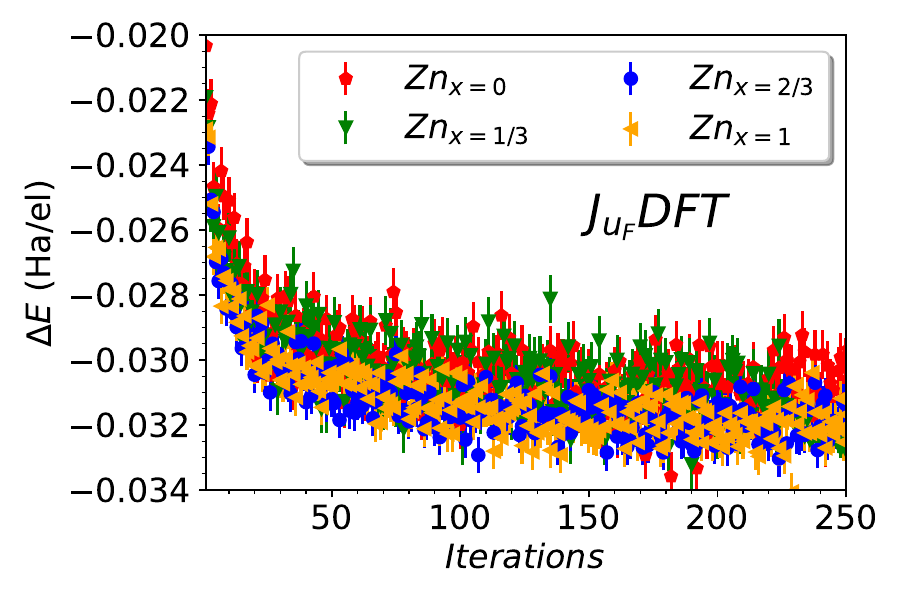} \\
        \includegraphics[scale=0.38]{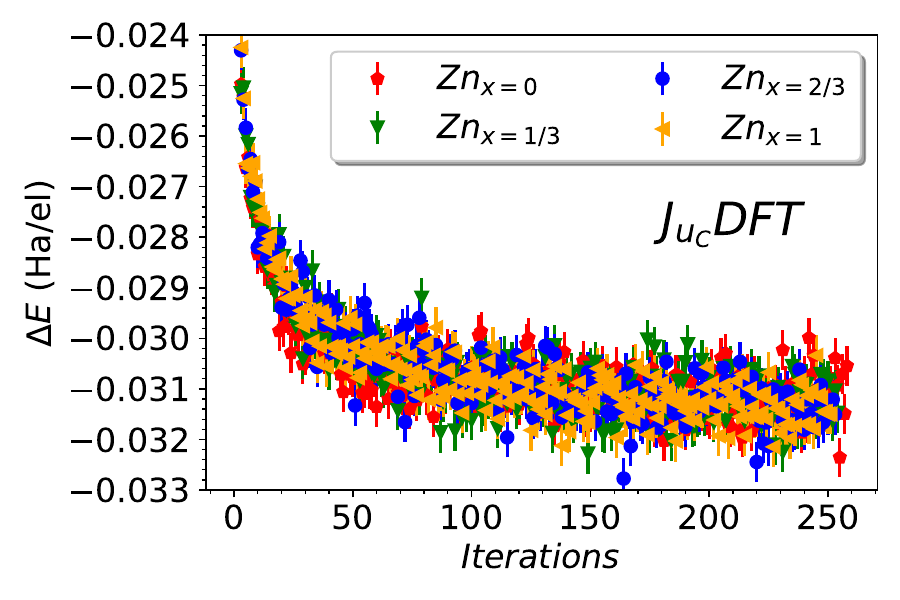} &
         \includegraphics[scale=0.38]{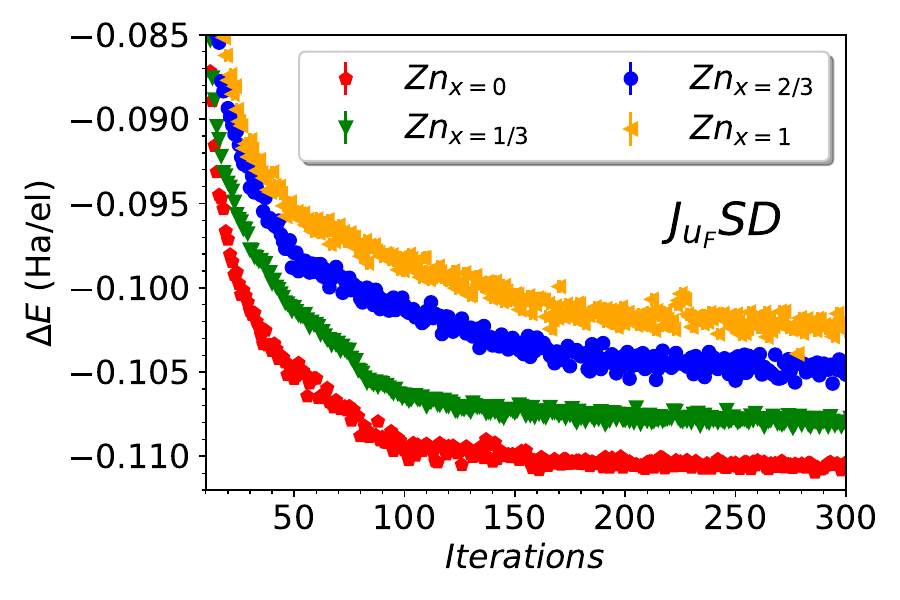} & 
          \includegraphics[scale=0.38]{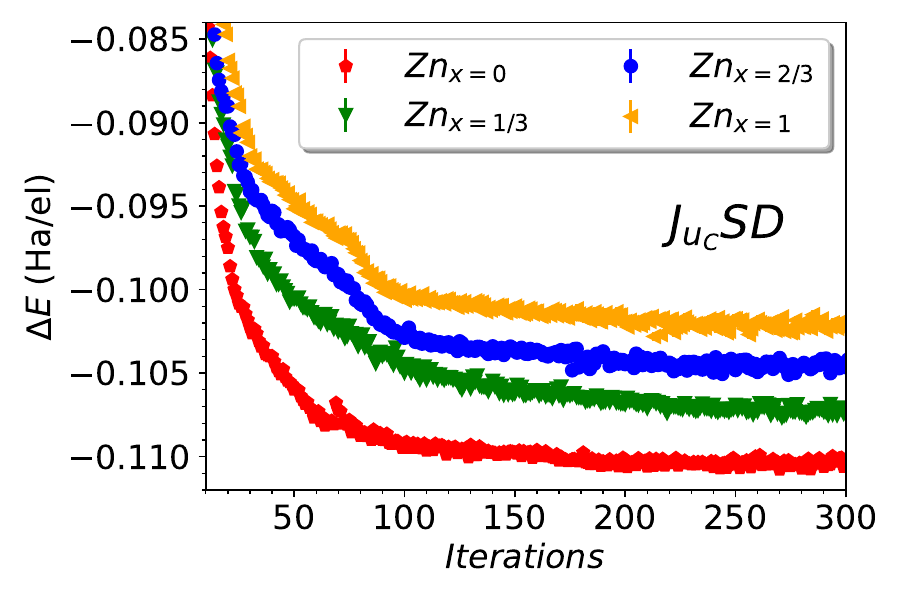} \\        
    \end{tabular}
    \caption{The difference between VMC optimization energy and DFT ($\Delta E$) as a function of the number of energy minimisation steps for $Zn_xCu_{4-x}O_6$ system obtained by three different wave functions of $J_{u_i}^{cof}DFT$, $J_{u_i}DFT$, $J_{u_i}SD$ where $i=C, F$ represent two different forms for two-body Jastrow term as explained in the text. Only the Jastrow coefficients and $a$, and $b$ parameters of two-body $u$ term were optimised in $J_{u_i}^{cof}DFT$ wave function. The Jastrow terms are fully optimised, both exponents and coefficients, in $J_{u_i}DFT$. Jastrow and Slater determinant are optimised simultaneously in $J_{u_i}SD$ wave function.  The results are obtained for $Zn_x$ concentrations of $x=0., 1/3, 2/3, 1.$. }
    \label{fig:OptWFC}
\end{figure*}

\subsection{Wavefunction optimization and VMC ground state energy}
We calculate the static and dynamic correlations between electrons as a function of concentration $Zn_x$. The correlation energy is defined as the difference between the exact ground state energy, which is obtained by the many-body wave function-based method VMC, and Hartree-Fock (HF). We also compare the correlation energy $E_c = E_{VMC} - E_{HF}$ with the energy difference between VMC and DFT $\Delta E = E_{VMC} - E_{DFT}$ where the correlation energy in LDA-DFT was implemented using the parameterization of a function based on the diffusion Monte Carlo energies of the uniform electron gas at high and low densities \cite{lda,Ceperley80}. We compare the variational energies obtained from two forms of homogeneous two-body Jastrow named $u_C$ and $u_F$, and discuss the importance of optimization of the determinant part of the trial WF. 

We systematically optimized our trial wave functions using the following steps:
\begin{enumerate}
    \item Step I: The Jastrow coefficients and $a$, and $b$ parameter of two-body $u$ term were optimised and the 
    wave function was called $J_{u_i}^{cof}DFT\text{-}WF$ where $i=C, F$ (Fig.~\ref{fig:OptWFC})
    \item Step II: The Jastrow Gaussian exponents plus all the variational parameters in step 1 were optimised and the WF was called $J_{u_i}DFT\text{-}WF$ 
    with $i=C,F$ (Fig.~\ref{fig:OptWFC})
    \item Step III: All variational parameters in both the Jastrow and Slater determinant were optimised and the WF was called $J_{u_i}SD\text{-}WF$ with 
    $i=C,F$ (Fig.~\ref{fig:OptWFC})
\end{enumerate}

We performed several hundred optimization steps to reach the accuracy of mHa/el. We found that following the above wave function optimization steps in order, meaning that the well-optimised wave function at each level should be used as the starting point for the next optimization step, is important to avoid local minima. Figure \ref{fig:OptWFC} summarizes the last few hundred iterations of the wave function optimization by energy minimization. The VMC energies at each optimization iteration are plotted with respect to DFT to present the energy gain at each wave function optimization level. The energy gain by $JSD\text{-}WF$ is almost three times larger than $JDFT\text{-}WF$ for all the studies systems and Jastrow factors. This indicates the importance of full optimization of the trial wave function to capture the accurate strongly correlated many-body effects.
\begin{figure}
    \centering
    \begin{tabular}{cc}
    \includegraphics[scale=0.275]{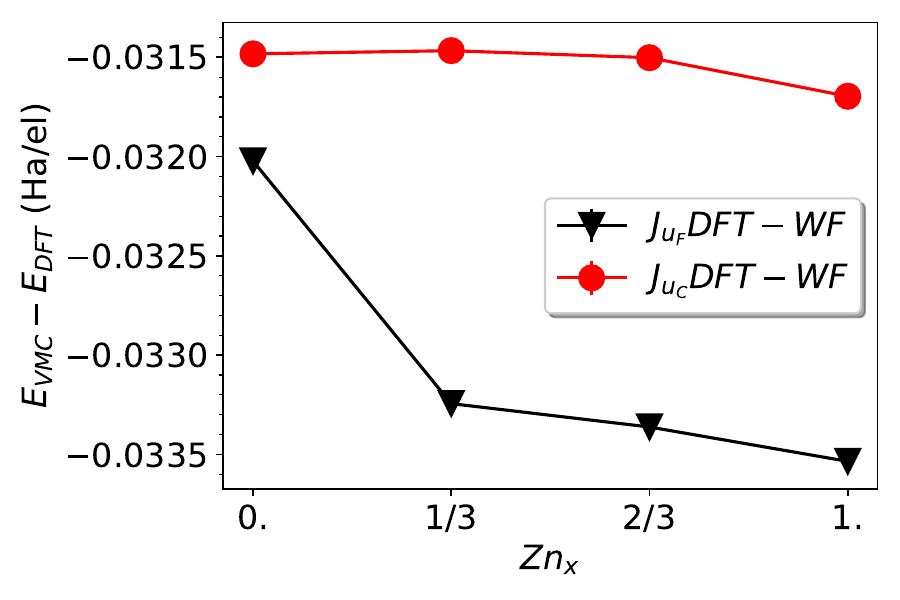} &
    \includegraphics[scale=0.275]{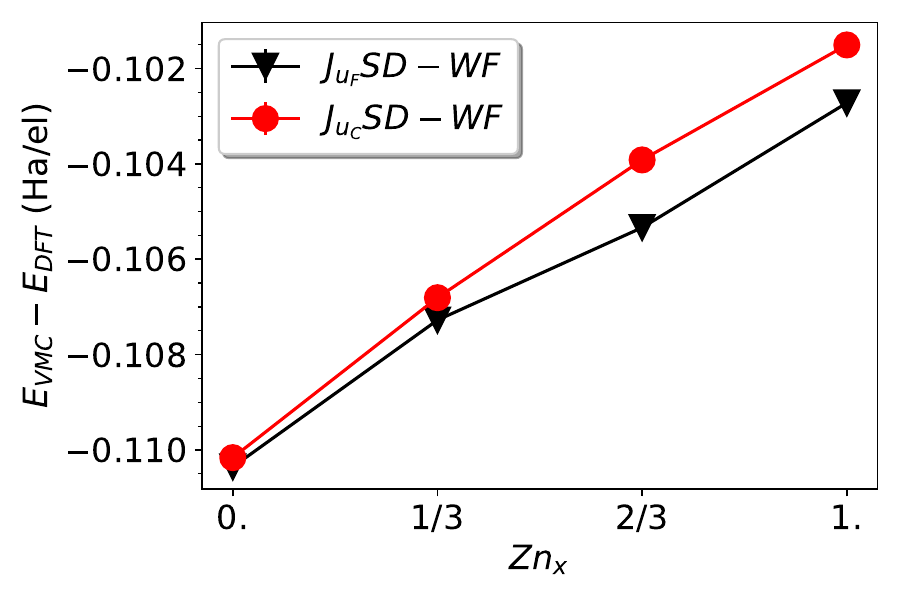}
    \end{tabular}
    \caption{The difference between VMC and DFT energies as a function of $Zn_x$ concentration in $Zn_xCu_{4-x}O_6$ obtained by $J_{u_i}DFT$ (left panel) and $J_{u_i}SD$ (right panel) wave functions where $i=F,C$. }
    \label{fig:vmc_DFT}
\end{figure}

The VMC energies obtained by the $JDFT$ and $JSD$ wavefunctions with two different Jastrow forms $u_C$ and $u_F$ are plotted in figure \ref{fig:vmc_DFT} as a function of the concentration of $Zn_x$. The two-body form $u_C(r)$ that we use is large and positive for $r=0$ and goes to zero as $r\rightarrow \infty$. Hence, it is useful whenever the atoms are far apart at a distance much larger than the $b$ parameter. This form produces a good size-consistent energies, which is approximately equal to the sum of the atomic contributions without significantly affecting the other parts of the wave function with an expensive optimization. The other form of the pair function $u_F(r)$, is especially convenient in atomic chemical bond separation. In principle, the VMC energies obtained from the two forms should agree with statistical noise, but we found that it depends on the trial wavefunction, as is explained below.

The difference between the VMC energy, which is obtained using two-body Jastrow terms of $u_C(r)$ and $u_F(r)$, and DFT (which is independent of the two-body term $u$) as a function of $Zn_x$ concentration is illustrated in Figure~\ref{fig:vmc_DFT}. The results obtained by $JDFT\text{-}WF$ show that the two-body function $u_C(r)$ produces VMC energies that are almost independent of $Zn_x$ concentration, while the VMC energies obtained by the $u_F(r)$ term decrease by $Zn_x$, as shown in the left panel of figure \ref{fig:vmc_DFT}. The same plot also indicates that there is a difference between the VMC-$u_C(r)$ and VMC-$u_F(r)$ energies that increases with $Zn_x$ concentration. We found that this difference becomes very small by fully optimizing the WF (Fig. \ref{fig:vmc_DFT} right panel). The difference between VMC-$u_C(r)$ and VMC-$u_F(r)$ energies obtained by $JSD\text{-}WF$ for $x=0, 1/3$ is almost zero and is $\sim 2(1)$ mHa/el for $x=2/3, 1$. The VMC energies calculated by the wave function $J_{u_C}SD$ and $J_{u_F}SD$ increase with the concentration of $Zn_x$. 

The difference between VMC energies obtained from homogeneous 2b-Jastrow $u_C$ and $u_F$, which is much more pronounced in $JDFT\text{-}WF$ than $JSD\text{-}WF$, as a function of $Zn_x$ concentration can be related to the behaviour of the used 2b-Jastrows in different densities. The 2b-Jastrow term in an inhomogeneous system wave function tries to make the charge density more uniform, than the one which is calculated from the mean-field Slater determinant, by reducing the charge density in large-density parts and enlarging the charge density in lower-density regions. However, the main reason for using the 2b-Jastrow in a many-body wave function is to improve the pair-correlation function, not the one-particle density, which is usually assumed to be well defined by DFT (or Hartree-Fock) \cite{BeccaSorella, Sorella2015}. Also, this difference indicates that in the $Zn_xCu_{4-x}O_6$ system the long-range correlations are important. Because two-body Jastrow $u_F$ decays much more slowly ($1/r$ tail) than two-body Jastrow $u_C$ term (exponential tail). The fact that we gain more energy with $u_F$ indicates that the long-range tails of $u_F$ are important to optimize the system. The results show that the electron-electron correlations become longer-ranged upon $Zn_x$ doping. It is interesting to notice that even after the full optimization of the wave function, the WF with the long-range Jastrow ($u_F\text{-}JSD$) is still slightly better than the one with the short-range one ($u_C\text{-}JSD$) for $Zn_x, x=2/3, 1$ systems. It has been observed that long-range correlations are important to stabilize spin-liquid phases in some spin-lattice Hamiltonians \cite{Zhou2017}.

\subsection{Correlation energy}

The Jastrow factor plays an important role in capturing dynamic many-body correlation effects, especially the two-body correlation function $u$, which is responsible for the most crucial correlation contribution introduced by the electron-electron interaction. The bosonic $u$ term reduces the probability of electron coalescence and therefore lessens the average value of the repulsive interaction. As was shown for uniform systems such as homogeneous electron gas \cite{Ceperley78}, when $u(r)$ is a positive and decreasing function of $r$ the probability of two electrons approaching each other decreases, resulting in a reduction in the energy of the electron-electron interaction. Hence, the kinetic term of the Hamiltonian is increased by introducing the Jastrow function and therefore the variational behavior of the $u(r)$ term is crucial to obtain the minimized energy \cite{Ceperley86}. As we discussed in the previous section the interplay between Jastrow terms and the single particle wave functions of the Slater determinant in finding the true ground state energy of inhomogeneous system $Zn_xCu_{4-x}O_6$ is complex and depends on the $Zn_x$ concentration.

\begin{table}
    \centering
    \resizebox{0.46\textwidth}{!}{%
    \begin{tabular}{|c|cccc|}
    \hline\hline
    WF & $E_c^{x=0}$ & $E_c^{x=1/3}$ & $E_c^{x=2/3}$ & $E_c^{x=1}$ \\
    \hline
    $J^{cof}_{u_F}DFT$ & $-0.78675(5)$ & $-0.79288(4)$ & $-0.79872(4)$ & $-0.80344(3)$ \\
    $J_{u_F}DFT$       & $-0.79875(3)$ & $-0.80469(3)$ & $-0.80949(4)$ & $-0.81432(3)$ \\
    $J_{u_F}SD$       & $-0.87708(2)$ & $-0.87873(2)$ & $-0.88147(2)$ & $-0.88351(2)$ \\
    \hline
    $J^{cof}_{u_C}DFT$ & $-0.78788(4)$ & $-0.79266(4)$ & $-0.79728(4)$ & $-0.80187(4)$ \\
    $J_{u_C}DFT$       & $-0.79821(3)$ & $-0.80291(3)$ & $-0.80763(3)$ & $-0.81248(3)$ \\
    $J_{u_C}SD$       & $-0.87689(2)$ & $-0.87825(2)$ & $-0.88005(2)$ & $-0.88228(2)$ \\
    \hline\hline
    \end{tabular}}
    \caption{The VMC correlation energy in Ha/el for $Zn_xCu_{4-x}O_6$ system with different values of $Zn_x$ concentration $x=0, 1/3, 2/3, 1$. Energies were calculated using three different wave functions of $J_{u_j}^{cof}DFT\text{-}WF$, $J_{u_j}DFT\text{-}WF$, and $J_{u_j}SD\text{-}WF$ with $j=F, C$ as explained in the text.}
    \label{tab:vmcenergy}
\end{table}

\begin{figure}
    \centering
    \begin{tabular}{cc}
    \includegraphics[scale=0.275]{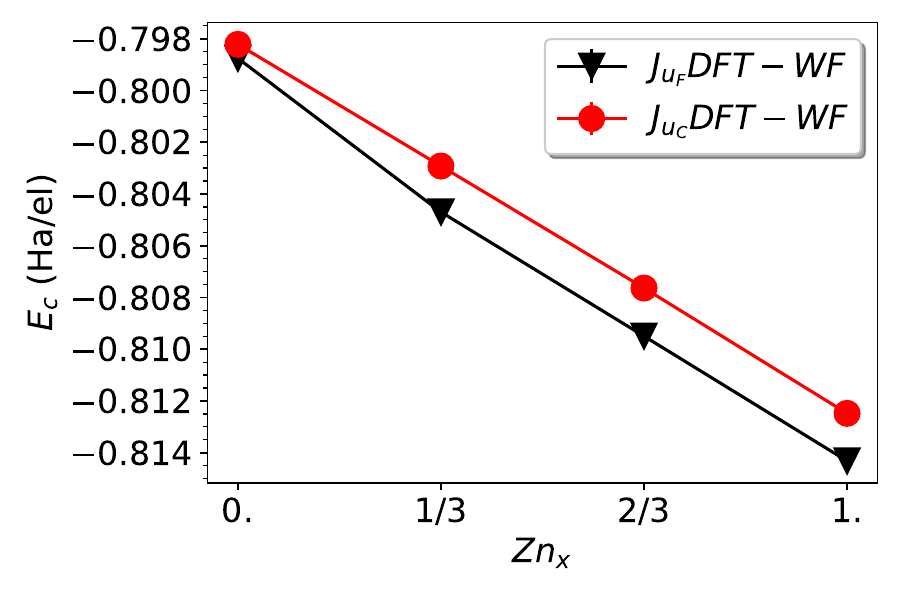} &
    \includegraphics[scale=0.275]{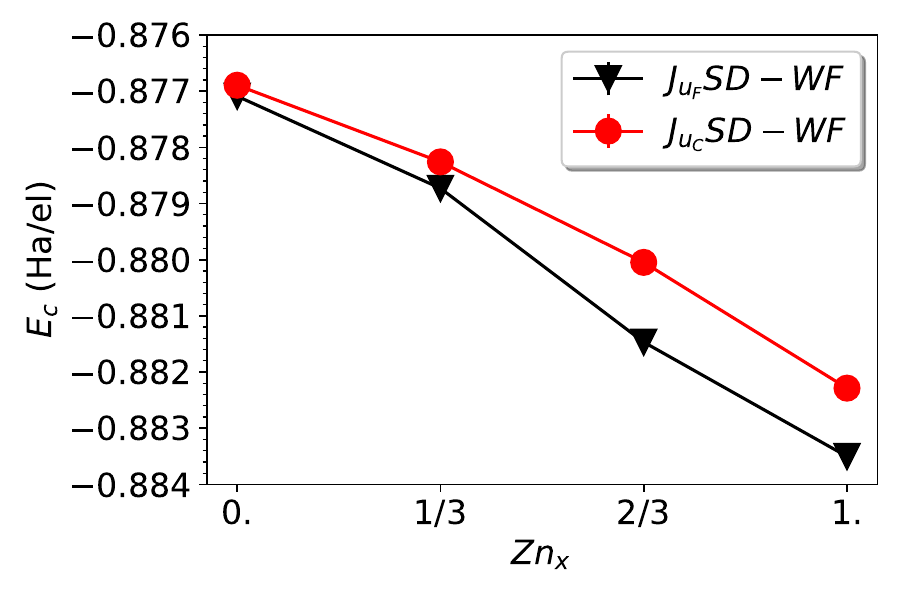}
    \end{tabular}
    \caption{The VMC correlation energies as a function of $Zn_x$ concentration in $Zn_xCu_{4-x}O_6$ system. (left panel) VMC energies were calculated using two forms of 2b-Jastrow $u_C$ and $u_F$ where in both wave functions the DFT-Slater determinant was not optimised. (right panel) The 2b-Jastrow term is the same as the left panel but the single-particle Slater determinant and the Jastrow were optimised simultaneously.}
    \label{fig:vmc_corr}
\end{figure}

The VMC correlation energy values for the concentration $Zn_x$ of $x=0., 1/3, 2/3, 1$ obtained by $J_{u_i}^{cof}DFT\text{-}WF$, $J_{u_i}DFT\text{-}WF$ and $J_{u_i}SD\text{-}WF$ are listed in the table\ref{tab:vmcenergy}. The results obtained by the wave functions $JDFT$ and $JSD$ indicate that the absolute value of the VMC correlation energy increases by increasing the concentration $Zn_x$ regardless of the form of the two-body Jastrow term $u$ (Fig.~\ref{fig:vmc_corr}). The correlation energy of the system with $x=0$ obtained by $u_C$ and $u_F$ is very close. Whereas, the absolute value of VMC $J_{u_F}SD$ correlation energies for the systems with $x=2/3, 1.$ is larger than VMC $J_{u_C}SD$ results. As we discussed in the previous section, the $u_F$ Coulomb $1/r$ tail decays more slowly than $u_C$. Hence, VMC $J_{u_F}SD$ results indicate that the long-range correlations become more important at concentrations $x=2/3, 1.0$ which are paramagnetic-like ground states even at temperatures near absolute zero \cite{Mendels2007}. 

The parent structure, $x$ = 0, of the compounds studied $Zn_xCu_{4-x}O_6$, named clinoatacamite, is a type of disordered $S=1/2$ pyrochlore with magnetic order that appears at 19 K \cite{Mendels2007,Bert2007}. The substitution of $Zn$ on the less Jahn-Teller distorted $Cu$ sites located between the kagome planes restores the three-fold symmetry of the lattice for $x > 1/3$ with a larger electron-electron correlation. The magnetic order gradually disappears as $x\rightarrow 1$ and for $x$=1 no spin freezing has been observed down to 50 mK \cite{Mendels2007}. Although all of our calculations are performed at zero temperature, our VMC energies show that there is a direct link between magnetic ordering and electron correlation. Our VMC-$JSD$ results show that the electronic correlation for the $x$=0 system with magnetic ordering at zero temperature is weaker than the $x$=1 compound with paramagnetic behaviour at nearly zero temperature \cite{Khuntia2020}. In addition, comparing $J_{u_C}SD$ and $J_{u_F}SD$ VMC correlation energies suggests that the long-range correlation between electrons appears by gradually disappearing the magnetic order from $x=0$ to $x=1$. Hence, one can conclude that the short-range exchange interaction, which is mainly responsible for the spin freezing of the $x=0$ compound at zero temperature, becomes weaker in comparison with the long-range correlation which is caused by increasing the $Zn_x$ concentration. 

\subsection{Effect of basis set on the correlation energy}
To determine the contribution of the static correlation energy, we calculated the VMC correlation energy of the $x\text{=}1$ system using two basis set sizes $10s8p6d2f$ and $10s8p8d4f$ (Table ~\ref{tab:basisset}). Three trial wave functions $J^{cof}DFT$, $JDFT$, and $JSD$ were used for each basis set size. The difference between VMC energy obtained by $10s8p6d2f$ and $10s8p8d4f$ basis sets with $J^{cof}DFT$, $JDFT$, and $JSD$ wave functions are $-9.50(4)$, $-6.96(3)$, and $-0.72(3)$ mHa/el, respectively. These results demonstrate the importance of full optimization of the trial wave function rather than relying on only the Jastrow optimization. 

\begin{table}[!ht]
    \centering
    \begin{tabular}{|cccc|}
    \hline\hline
    Basis set & $J^{cof}DFT$ & $JDFT$ & $JSD$ \\
    \hline
    $10s8p6d2f$ & $-0.80344(3)$& $-0.81432(3)$& $-0.88351(2)$ \\
    $10s8p8d4f$ & $-0.81294(2)$& $-0.82128(2)$& $-0.88423(2)$ \\
    \hline\hline
    \end{tabular}
    \caption{VMC correlation energy of $x=1$ system obtained by three wave functions of $J^{cof}DFT$, $JDFT$, and $JSD$, and two basis set size. Energies are in Ha/el.}
    \label{tab:basisset}
\end{table}

\subsection{Electron pairing energy}\label{AGPWF}
We performed extensive wave function optimization using $JAGP$ ansatz with $u_F$ two-body Jastrow form for $x\text{=}1$ system. We found that the VMC energy difference between the $JSD$ and $JAGP$ wave functions is within the error bar. Furthermore, we calculated the RVB pairing energy in the $CuO_2$ plane with the same geometry as it is adopted in the compound $x\text{=}1$.  The $Cu$ atoms in the $ZnCu_3(OH)_6Cl_2$ trigonal crystal structure (Fig.~\ref{fig:herb}) are located in a kagome lattice that is a two-dimensional ($2d$) network of corner-sharing triangles. 
\begin{figure}
    \centering
    \includegraphics[scale=0.5]{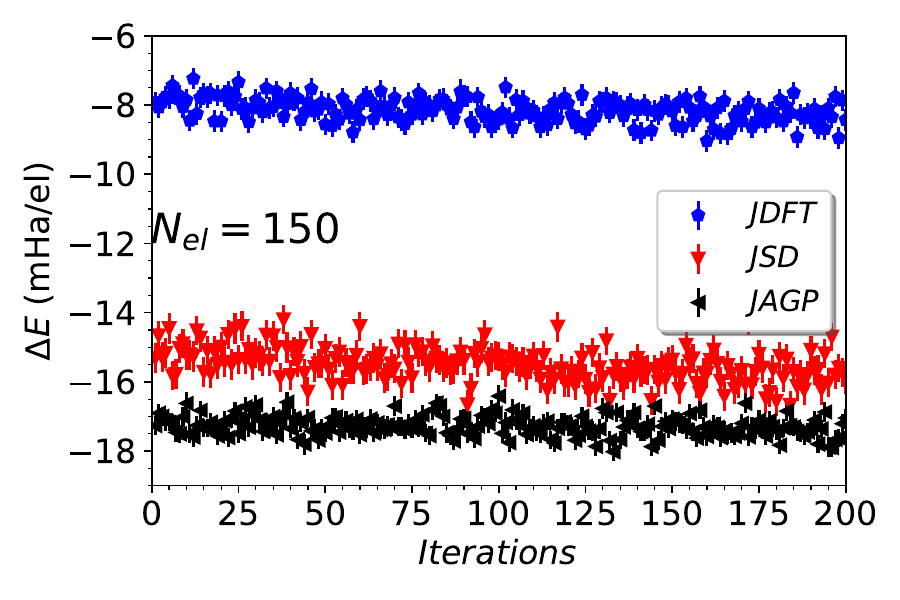}
    \includegraphics[scale=0.5]{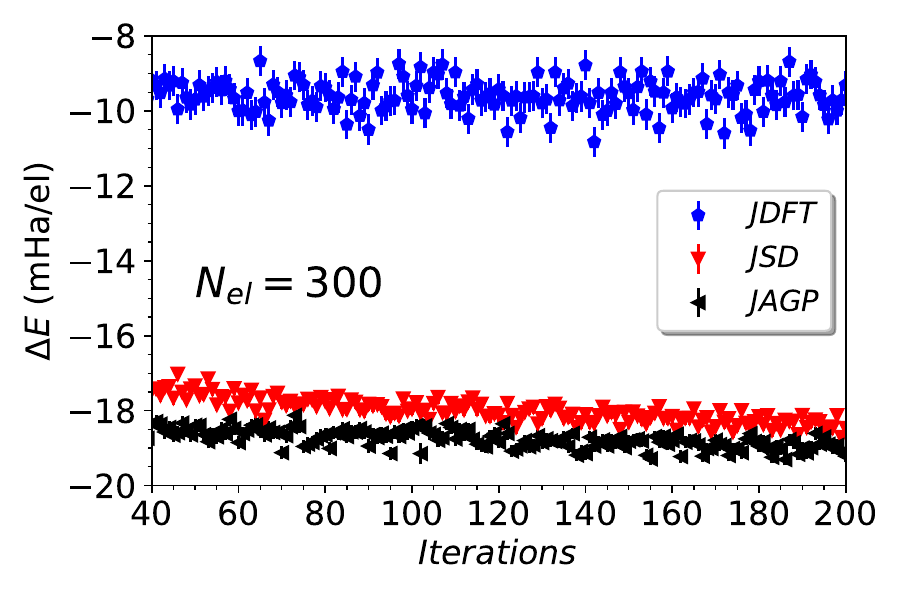}
    \caption{The difference between VMC energy minimization and DFT ($\Delta E$) as a function of the number of optimization steps for the two-dimensional $CuO_2$ layer. Two system sizes with the number of electrons per simulation cell of $N_{el}=150$ (top panel) and $N_{el}=300$ (bottom panel) were used. The VMC energies were obtained with three trial wave functions of $JDFT$, $JSD$, and $JAGP$ as explained in the text.}
    \label{fig:optagp}
\end{figure}

We studied two system sizes of $2d\text{-}CuO_2$ with the number of electrons per simulation cell of $N_{el}=150$ and $300$. We performed several hundreds of wave function optimization steps using $JDFT$, $JSD$, and $JAGP$ wave functions. Figure \ref{fig:optagp} illustrates the energy difference between VMC and DFT within the last couple of hundreds of WF optimization steps. As we observed earlier, full wave function optimization plays a crucial role in improving the correlation energy. Comparison of $JDFT$ and $JSD$ variational energies clearly shows that the $JSD$ VMC energy gain is almost two times greater than $JDFT$.
\begin{table}
    \centering
    \begin{tabular}{|cccc|}
    \hline\hline
    $N_{el}$ & $JDFT$ & $JSD$ & $JAGP$ \\
    \hline
    $150$ & $-8.46496(3)$& $-8.47384(3)$& $-8.47609(3)$ \\
    $300$ & $-8.46454(2)$& $-8.47322(2)$& $-8.47384(2)$ \\
    \hline\hline
    \end{tabular}
    \caption{VMC energy of two dimensional $CuO_2$ layer obtained by $JDFT$, $JSD$, and $JAGP$ wave function for two different system sizes. Energies are in Ha/el.}
    \label{tab:agpvmc}
\end{table}

Our VMC ground state energy for $2d\text{-}CuO_2$ obtained using the wave functions $JDFT$, $JSD$, and $JAGP$ are listed in table~ \ref{tab:agpvmc}. The difference between the $JAGP$ and $JSD$ VMC energies of systems with $N_{el}=150$ and $N_{el}=300$ is $-2.25(4)$ and $-0.62(3)$ mHa/el, respectively. This suggests that the RVB energy gain in the $2d-CuO_2$ plane is short-ranged and remains within a relatively small resonance length of a few atomic units. The finite-size (FS) errors play a crucial role in the accurate description of RVB pairing energy as the difference between $JSD$ and $JAGP$ VMC energies is small. Therefore, the description of RVB pairing energy at the thermodynamic limit requires QMC simulations for larger system sizes and extrapolation techniques. 

Copper oxides, especially $CuO_2$ planes, play an important role in the study of magnetism and high-temperature superconductivity in cuprates \cite{Kivelson2003,PALee2006}. Although the precise mechanism of high-temperature superconductivity in cuprates is still the subject of intense research, our RVB results indicate that electron pairing occurs at the $CuO_2$ planes. We have shown before that the existence of $d$-wave pairing symmetry in these materials predicts a high degree of anisotropy in the superconducting gap \cite{Marchi2011}. Since the $CuO_2$ plane exhibits an antiferromagnetic order, the singlet pairing mechanism is related to the antiferromagnetic spin fluctuations, leading to long-range magnetic order. Anderson \cite{Anderson1987} proposed for the first time that $CuO_2$ planes in cuprates could establish a spin-liquid phase with pairs of spins forming singlets that resonate across different bonds. Our results suggest that upon doping the magnetic ordering is suppressed by long-range correlation which results in the stability of a paramagnetic-like spin-liquid state. 

\section{Conclusion}\label{conclude}
We used the VMC method to study the ground state electronic structure of $Zn_xCu_{4-x}O_6$. We calculated the variational correlation energy between electrons using two different trial wave functions named $JDFT$ and $JSD$ as a function of the concentration $Zn_x$. The Slater determinant obtained by LDA-DFT was not optimized in $JDFT$ but in $JSD$. A significant improvement in static correlation energy was observed by optimizing the Slater determinant. We used two different forms, labeled $u=C$ and $u=F$, for the homogeneous two-body part of the Jastrow function and compared the results. We found that the VMC correlation energy obtained by the wave functions $JDFT$ and $JSD$ increases with the concentration $Zn_x$. 

We calculated the pairing energy using the JAGP wave function. We found that the pairing energy in $ZnCu_3O_6$ is negligible. However, the VMC calculations for the $CuO_2$ layer with $N_{el}=150$ and $300$ give the pairing energy of -2.25(4) and -0.62(3) mHa/el, respectively.  
\section{Acknowledgement}
S.A. acknowledges support from the Leverhulme Trust under the grant agreement RPG-2023-253. The authors gratefully acknowledge the computing time provided to them on the high-performance computers Noctua2 at the NHR Center in Paderborn (PC2).
\bibliography{main}
\end{document}